\def\expec#1{\langle#1\rangle}
\def\bexpec#1{\left\langle#1\right\rangle}

\def\etal{{\frenchspacing\it et al.}}
\def\ie{{\frenchspacing\it i.e.}}
\def\eg{{\frenchspacing\it e.g.}}
\def\etc{{\frenchspacing\it etc.}}
\def\rms{{\frenchspacing r.m.s.}}

\def\rf#1;#2;#3;#4;#5 {\par#1, {\it #3} {\bf #4}, #5 (#2). \par}

\def\beq#1{\begin{equation}\label{#1}}
\def\eeq{\end{equation}}
\def\beqa#1{\begin{eqnarray}\label{#1}}
\def\eeqa{\end{eqnarray}}
\def\eq#1{equation~(\ref{#1})}
\def\Eq#1{Equation~(\ref{#1})}

\def\fig#1{Figure~\ref{#1}}
\def\Fig#1{Figure~\ref{#1}}

\def\simpropto{\,{\lower3.0pt\hbox{$\propto$}\atop\raise1.0pt\hbox{$\sim$}}\,}

\def\C{{\bf C}}
\def\E{{\bf E}}
\def\F{{\bf F}}
\def\Ft{{\tilde{\bf F}}}

\def\I{{\bf I}}
\def\J{{\bf J}}
\def\L{{\bf L}}
\def\M{{\bf M}}
\def\N{{\bf N}}

\def\P{{\bf P}}
\def\Pt{{\tilde{\bf P}}}
\def\PP{{\bf\Pi}}
\def\R{{\bf R}}

\def\U{{\bf U}}
\def\V{{\bf V}}
\def\V{{\bf V}}
\def\W{{\bf W}}
\def\Wt{{\tilde{\bf W}}}
\def\Win{\W}
\def\X{{\bf X}}
\def\Y{{\bf Y}}
\def\Z{{\bf Z}}
\def\0{{\bf 0}}
\def\m{{\bf m}}
\def\a{{\bf a}}
\def\x{{\bf x}}
\def\y{{\bf y}}
\def\z{{\bf z}}

\def\xt{\tilde{\bf x}}
\def\yt{{\tilde{\bf y}}}
\def\Ct{{\tilde{\bf C}}}
\def\Ctt{{\tilde{\bf C}}_*}
\def\rh{{\widehat{\bf r}}}

\def\l{\ell}
\def\lmax{\l_{max}}
\def\c{{\bf c}}
\def\ct{\tilde{\bf c}}
\def\Cl{C_\l}
\def\Ch{\widehat{C}}
\def\D{D}
\def\Dh{\widehat{\D}}

\def\tr{\hbox{tr}\>}
\def\b{b}
\def\Ell{L}
\def\norm{N}
\def\Qrmsps{Q_{rms,ps}}

\def\mK{{\rm \mu K}}

\documentstyle[prd,twocolumn,aps,epsf,rotate]{revtex}
\begin{document}




\preprint{IASSNS-AST 96/63}

\title{How to measure CMB power spectra without losing information}

\author{Max Tegmark\thanks{Hubble Fellow.}}

\address{Institute for Advanced Study, Princeton, NJ 08540; max@ias.edu}

\address{Max-Planck-Institut f\"ur Astrophysik,  
Karl-Schwarzschild-Str. 1, D-85740 Garching}

\date{Submitted November 21, 1996; accepted February 26, 1997}

\maketitle

\begin{abstract}
\noindent{\bf Abstract:} 
A new method for estimating the angular power spectrum 
$C_\l$ from cosmic microwave background (CMB) maps is presented, 
which has the following desirable properties:\\
(1) It is unbeatable in the sense that no other method can measure
$C_\l$ with smaller error bars.\\
(2) It is quadratic, which makes the statistical properties of
the measurements easy to compute and use for estimation 
of cosmological parameters.\\
(3) It is computationally faster than rival high-precision
methods such as the nonlinear maximum-likelihood technique,
with the crucial steps scaling as $n^2$ rather than $n^3$,
where $n$ is the number of map pixels.\\
(4) It is applicable to any survey geometry whatsoever, with arbitrary
regions masked out and arbitrary noise behavior.\\
(5) It is not a ``black-box" method, but quite simple to understand
intuitively: it corresponds to a high-pass filtering
and edge softening of the original map followed by a straight 
expansion in truncated spherical-harmonics.\\
It is argued that this method is computationally feasible
even for future high-resolution CMB experiments
with $n\sim 10^6-10^7$. It is shown that
$C_\l$ computed with this method is useful not merely for 
graphical presentation purposes, but also as an 
intermediate (and arguably necessary) step in the data analysis 
pipeline, reducing the
data set to a more manageable size before 
the final step of constraining Gaussian
cosmological models and parameters --- while retaining 
all the cosmological information that was present in
the original map.
\end{abstract}


\makeatletter
\global\@specialpagefalse
\def\@oddfoot{
\ifnum\c@page>1
  \reset@font\rm\hfill \thepage\hfill
\fi
\ifnum\c@page=1
  {\sl Accepted for publication in Phys. Rev. D.
Available in color from 
h t t p://www.sns.ias.edu/$\tilde{~}$max/cl.html} \hfill\\
\fi
} \let\@evenfoot\@oddfoot
\makeatother



\section{INTRODUCTION}

The angular power spectrum $\Cl$ of the fluctuations in the 
cosmic microwave background (CMB) is a gold mine of cosmological
information. Since it depends on virtually all classical cosmological 
parameters (the Hubble parameter $h$, the density parameter 
$\Omega$, the cosmological constant $\Lambda$, {\etc}),
an accurate
measurement of $\Cl$ would amount to an accurate measurement 
of most of these parameters 
\cite{JKKS1,JKKS2,BET}.
In the last few years, the angular power spectrum has emerged as the
standard way of presenting experimental results in the literature, 
replacing other fluctuation measures such as the correlation function and the 
Gaussian autocorrelation function amplitude
(as described in {\eg} \cite{WSS}).
There are several reasons for this:
\begin{enumerate}
\item 
The Boltzmann equation is  diagonal in the Fourier (multipole) domain
rather than in real space, so the features of the power spectrum can 
be given a direct and intuitive physical interpretation 
(see {\eg} \cite{HSS}).
\item 
A plot of power-spectrum estimates allows experiments to be 
compared in a model-independent way, as opposed to, say, 
parameter estimates and exclusion plots
that are only valid within the framework of
particular cosmological models.
\item 
For Gaussian models, power spectrum estimation constitutes a
useful (and arguable necessary) data compression trick for making the analysis 
of future megapixel sky maps feasible in practice.
\end{enumerate}
In what follows, we will pay considerable attention to 
the the third point, since there are at present no unbeatable 
methods available that are computationally feasible when $n$, 
the number of map pixels, is very large.
The CPU time needed for applying the 
maximum-likelihood method directly to a map scales as $n^3$, since
it involves computing determinants of $n\times n$ (non-sparse) covariance
matrices, and the Karhunen-Lo\`eve data compression method
(see \cite{Bond,Bunn+Sugiyama,Bunn,karhunen} and references therein) 
unfortunately requires the diagonalization of an
$n\times n$ matrix, which also scales as $n^3$.
Such a brute-force approach has so far only been implemented up to  
$n\sim 4000$ \cite{karhunen,brute,Hinshaw}, and it currently appears unfeasible to push it much 
beyond $n=10^4$. 
In contrast, the upcoming satellite missions 
{\it MAP} and {\it Planck} will
have $n$ in the range $10^6-10^7$. 
The data-compression aspect of power spectrum estimation is 
illustrated in Figure 1: if the power spectrum retains
(in a distilled form)
all the cosmological information that was present in the map, then
the computationally unfeasible step of estimating the parameters
directly from the map can be split in to two feasible steps,
giving exactly the same answer and and the same error bars. 
This is completely analogous to the way in which map-making
is an intermediate step, and as was recently shown \cite{mapmaking},
there are indeed map-making methods that destroy no 
cosmological information at all.

In this paper, 
we will derive a new method for estimating $\Cl$ from maps that has the
following desirable properties:
\begin{enumerate}
\item It is {\it unbeatable} in the sense that no other method can give smaller
error bars on $C_\l$ or on any cosmological parameters
upon which $C_\l$ depends.
\item It is {\it quadratic}, which means that the statistical properties 
of the estimates are easy to compute.
\item It is {\it faster} than the maximum-likelihood method and 
the eigenmode method \cite{window}, with
the required CPU time for the crucial steps scaling as $n^2$ rather than $n^3$.
\item It is transparent and easy to understand intuitively.
\end{enumerate}
The rest of this paper is organized as follows. In 
Section 2, we discuss in more detail how to assess the merits of a power spectrum 
estimation method, and derive a simple test for determining whether it 
is unbeatable in the above sense. 
In Section 3, we derive the new method and prove that it is in fact unbeatable in
the sense of Section 2. 
In Section 4, we explore its properties, illustrated with an application to
the 4 year COBE/DMR data.
In section 5, we discuss how to use this method in the 
analysis of a future megapixel map, both for graphically presenting the data
and for measuring cosmological parameters. Finally, we summarize our conclusions in 
Section 6.

\section{HOW TO ASSESS HOW GOOD A METHOD IS}

Above we listed three uses for power spectrum estimation methods.
Especially for the third use, as a data compression technique,
we clearly want a method to have the following properties:
\begin{enumerate}
\item
It should be computationally feasible in practice.
\item 
It should produce estimates of $\Cl$ whose statistical properties
are well enough understood to make them useful for parameter estimation
and model testing. 
\item 
It should destroy as little information as possible.
\end{enumerate}
We will now elaborate on the third of these criteria, and return to the 
other two further on.

\subsection{The notion of a lossless method}

The Fisher information matrix formalism (see \cite{karhunen} for a comprehensive review) 
offers a simple and a useful way of diagnosing the methods corresponding
to the various boxes in Figure 1, to measure how much information
they destroy. Given any set of cosmological parameters of interest
($h$, $\Omega$, {\etc}), 
their Fisher matrix $\F$ gives the smallest error bars with which the parameters
can possibly be measured from a given data set. 
$\F^{-1}$ can, crudely speaking, be thought of as the 
best possible covariance matrix for the measurement errors on the parameters.
For instance, the 
Cramer-Rao inequality shows that no unbiased method whatsoever 
can measure the $i^{th}$ parameter with error bars (standard
deviation) less than $1/\sqrt{\F_{ii}}$.
If the other parameters are not known but estimated
from the data as well, the minimum standard deviation
rises to $(\F^{-1})_{ii}^{1/2}$.

By computing the Fisher matrix separately from each of the intermediate data
sets in Figure 1, we can thus track the flow of information
down the data pipeline and check for leaks.
For instance, if the Fisher matrix computed from the raw time-ordered
data (TOD) is identical to that computed from the map, then 
the map-making method (denoted $\W$ in the figure) is {\it lossless}
in the sense that no information about these parameters has been lost in
the map-making process. 
It was recently shown \cite{mapmaking} that some of the popular map-making methods
from the literature are lossless whereas others are not.
The advantage of making a lossless map is
that this reduces the data set to a more manageable size before the more
complicated nonlinear data analysis step (the final likelihood analysis).
We will see that the angular power spectrum plays 
quite an analogous role, allowing us to subject the map to a second 
data compression step before commencing the final
parameter estimation step, and we can clearly diagnose it
in exactly the same way.
Let us make the following 
definition, which is applicable to any data compression method whatsoever
(to any procedure that reduces a larger data set into a smaller one):
\begin{itemize}
\item 
{\it A data compression method is said to be {\it lossless}
if any set of cosmological parameters can be measured 
just as accurately from the compressed data set as from  
the original data set.}
\end{itemize}

\subsection{Lossless or not? A simple test}

Unfortunately, this definition is not particularly 
useful for diagnosing a method in practice, 
since it involves computing the Fisher matrices for a large or infinite
number of parameter sets. 
Fortunately, this is equivalent to a much simpler test, as we will now show.

If the probability distribution $f$ for the data set $\x$ (the pixels
temperatures in a sky map) depends on some parameters
$\lambda_1,\lambda_2,...$, then the Fisher information matrix for these parameters
is defined as \cite{karhunen}
\beq{FisherDefEq}
\F^{\lambda}_{ij}\equiv
-\bexpec{{\partial^2\over\partial\lambda_i\partial\lambda_j}\ln f}.
\eeq
Since $f$ is a probability distribution over $\x$, 
$\int f(\x;\lambda_1,\lambda_2,...)d^nx =1$ for any choice of the 
parameter vector $\{\lambda_i\}$. 
Differentiating this identity, we obtain
\beq{VanishingMeanEq}
\bexpec{{\partial\over\partial\lambda_i}\ln f}
= \int {\partial\ln f\over\partial\lambda_i} f d^nx 
= {\partial\over\partial\lambda_i}\int f d^nx = 0.
\eeq
Using this result and the chain rule, we find that if the parameter 
set $\{\lambda_i\}$
depends on some other parameter set $\{\theta_i\}$, then 
the Fisher matrix for these new parameters is given by 
\beq{FisherTransformationEq}
\F^{\theta} = \J^t\F^{\lambda}\J,
\eeq
where the Jacobian matrix
\beq{JdefEq}
\J_{ij}\equiv {\partial\lambda_i\over\partial\theta_j}.
\eeq
Note that this simple transformation rule holds regardless of whether the
probability distribution is Gaussian or not.
  
If the CMB fluctuations $\x$ are Gaussian and isotropic, 
then we know that their probability
distribution is entirely determined by the power spectrum. This means that
if we choose the parameters $\lambda_i$ to be the power spectrum 
coefficients $\C_\l$, the Fisher matrix for any cosmological
parameters whatsoever can be computed directly from 
$\F^C$, the Fisher matrix 
for the power spectrum itself:
\beq{FisherTransformationEq2}
\F^{\theta} = \J^t\F^{C}\J,
\eeq
where 
\beq{JdefEq2}
\J_{\l i}\equiv {\partial\Cl\over\partial\theta_i}.
\eeq
In other words, there is no need to compute and compare large numbers of Fisher
matrices for various parameter combinations, since they can all be computed
directly from $\F^C$.
Here and throughout, we will let $\Ch_\l$ denote
estimates of the true angular power spectrum 
$\Cl$, so the estimates are {\it unbiased} if they satisfy
\beq{UnbiasedEq}
\expec{\Ch_\l} = C_\l.
\eeq
With this notation, we can summarize this section as follows.
To test if a power spectrum estimation method is lossless, 
we simply compute the covariance matrix 
\beq{CovDefEq}
\V_{\l\l'}\equiv\expec{\Ch_\l\Ch_{\l'}} - \expec{\Ch_\l}\expec{\Ch_{\l'}}
\eeq
and check if it equals the inverse of the 
Fisher matrix $\F^C$ of \eq{FCeq} below.

\subsection{The power spectrum Fisher matrix}

Let us now evaluate this important matrix $\F^C$.
Using the addition theorem for spherical harmonics
gives the well-known correlation function formula
\beq{CsumEq}
\C\equiv\expec{\x\x^t} = \N + \sum_\l \P^\l\C_\l,
\eeq
where $\N$ denotes the noise covariance matrix and
the matrices $\P^\l$ are defined as
\beq{PdefEq}
\P^\l_{ij} \equiv {2\l+1\over 4\pi} P_\l(\rh_i\cdot\rh_j).
\eeq
Here the $P_\l$ denote Legendre polynomials and $\rh_i$ is a unit vector
pointing in the direction of pixel $i$. 
Thus $\partial\C/\partial\Cl=\P^\l$, and 
equation (15) of \cite{karhunen} 
(which gives the Fisher matrix for a general 
multivariate Gaussian probability distribution)
yields 
\beq{FCeq}
\F^C_{\l\l'} = {1\over 2}\tr\left[
\C^{-1}{\partial\C\over\partial C_\l} 
\C^{-1}{\partial\C\over\partial C_\l'}\right]
= {1\over 2}\tr\C^{-1}\P^\l\C^{-1}\P^{\l'}.
\eeq

Are there any lossless methods? Below we will answer this question 
affirmatively. 

\section{THE OPTIMAL METHOD}

In this section, we will derive the above-mentioned lossless method.

\subsection{A first guess: the ML-method}
\label{FirstGuessSec}

In many cases (including some mapmaking algorithms \cite{mapmaking}), 
the maximum-likelihood
(ML) method turns out to be lossless, so one might guess that this would
be the case here as well. Indeed, this approach to power-spectrum estimation  
has been applied to the 4 year COBE DMR data \cite{Hinshaw,Bunn+White}. 
Unfortunately, the ML-estimates $\Ch_\l$ turn out
to depend on the data set $\x$ in a highly nonlinear way, which gives the 
ML-estimates two undesirable properties:
\begin{enumerate}
\item They must be found by numerically solving a system of 
nonlinear equations, which is time-consuming. 
\item The probability distributions for these estimates are virtually 
hopeless to compute analytically, which makes it difficult to use 
the ML-power spectrum estimates in the last step of the data pipeline, in 
a likelihood analysis to determine cosmological parameters.
\end{enumerate}
For these reasons, it would be a pleasant surprise if the ML-method 
turned out {\it not} to be lossless, but inferior to some simpler
power spectrum estimation technique.

\subsection{A second guess: quadratic methods}

Fortunately, as we will see below, there are indeed considerably simpler 
estimates of the power spectrum that are lossless --- specifically, 
quadratic ones.
By a {\it quadratic estimator}, we mean one that is a quadratic function 
of the pixels, taking the form 
\beq{GeneralQuadraticEq}
\Ch_\l = \x^t\E^\l\x - \b_\l
\eeq
for some symmetric matrix $\E^\l$ and some constant $\b_\l$.
Before embarking on detailed calculations, let us give
a more intuitive argument for why we might expect
the best method to be quadratic.
It is easy to see that the entire data set $\x$ can be recovered from the
set of all pair products $x_ix_j$, apart from an uninteresting overall 
sign ambiguity: given the matrix $\x\x^t$, we simply
compute $x_i=\pm(\x\x^t)_{ii}^{1/2}$ and then fix all signs except one using 
the off-diagonal terms. Since the overall sign is irrelevant, the
data set consisting of the $n^2$ numbers in $\x\x^t$ therefore contains all the
cosmological information that the
$n$ numbers in $\x$ did.
Our quadratic power estimator
$\Ch_\l = \sum_{ij}\E_{ij}(\x\x^t)_{ij} -\b_\l$ is simply
a {\it linear} function of
these pair products. 
\Eq{CsumEq} is telling us that  
\beq{PairExpecEq}
\expec{x_i x_j} = \N_{ij} + \sum_\l \P_{ij}^\l\C_\l,
\eeq
{\ie}, that these pair products are on average just
{\it linear} 
combinations of the coefficients $C_\l$ that we want to measure,
so by analogy with the mapmaking results of \cite{mapmaking}, we might guess that
since the problem is linear, the best solution should be linear,
so that there exists an estimator of the form of \eq{GeneralQuadraticEq}
that is lossless.

Encouraged by this, we will now derive the the 
best method in the quadratic family. After that, we will give
a proof showing that this method is lossless, {\ie}, that
no other (more nonlinear) method can possibly do any better.

\subsection{The best quadratic method...}

Let us now find the quadratic power spectrum estimators that give the smallest
error bars. Substituting \eq{CsumEq} into \eq{GeneralQuadraticEq} and chosing 
\beq{BiasEq}
\b_\l=\tr\N\E^\l
\eeq
to make the estimate unbiased, we obtain
\beq{EstExpecEq}
\expec{\Ch_\l} = \sum_{\l'} \Win_{\l\l'}\C_{\l'},
\eeq
where the {\it Window function} is given by 
\beq{WindowDefEq}
\Win_{\l\l'}\equiv\tr\P^{\l'}\E^{\l}.
\eeq
Let us find the estimate of $\Cl$ with minimal variance subject to the normalization
constraint that $\Win_{\l\l}=1$.
Since we are assuming Gaussianity, the covariance matrix of 
\eq{CovDefEq} is given by
\beq{ClumsyCovEq}
\V_{\l\l'} = \sum_{ijkl} \left[\C_{ik}\C_{jl}+\C_{il}\C_{jk}\right]\E^{\l}_{ij}\E^{\l'}_{kl},
\eeq
so we want to find the $\E^\l$ that minimizes $\V_{\l\l}$ subject to this constraint.
The analogous problem for Galaxy surveys was recently solved by 
Hamilton \cite{H97a},
and we will follow his
notation and let a Greek index denote a pair of Latin indices, with
$\alpha=(i,j)$ and $\beta=(k,l)$. With this notation, 
$\E^\l$ and $\P^\l$ change from $n\times n$ matrices to 
$n^2$-dimensional vectors (or, since they are symmetric, to 
$n(n+1)/2$-dimensional vectors if we restrict ourselves to counting 
each pixel pair only once --- say, $j\ge i$). Defining the
matrix
\beq{HugeMdefEq}
\M_{\alpha\beta} \equiv\left[\C_{ik}\C_{jl}+\C_{il}\C_{jk}\right],
\eeq
our problem reduces to simply minimizing 
$\sum_{\alpha\beta}\M_{\alpha\beta}\E^\l_\alpha\E^\l_\beta$ 
subject to the constraint $\sum_\alpha\P^\l_\alpha \E_\alpha=1$.
Introducing a Lagrange multiplier just as in \cite{H97a}, we find the solution
to be
\beq{UselessSolutionEq}
\E^\l_\alpha \propto (\M^{-1})_{\alpha\beta}\P^\l_\beta.
\eeq

\subsection{...is in fact both simple...}

Unfortunately, \eq{UselessSolutionEq} is not a very useful result for our application, 
since the matrix that needs to be inverted is enormous, 
with dimensions $[n(n+1)/2]\times[n(n+1)/2]$ when eliminating the
redundant rows and columns corresponding to double-counted pixel pairs.
For this reason, Hamilton proceeds \cite{H97b} to provide an approximate method for 
solving this equation by means of a 
perturbation series expansion.
Fortunately, the giant matrix $\M^{-1}$ can be rewritten in a much simpler
form using some algebraic tricks.\footnote{
Note that for the non-Gaussian case, which is relevant 
{\eg} for non-linear clustering in Galaxy surveys,
this trick does not work, in which case the above-mentioned
perturbation series expansion of Hamilton is the only
approach presently available.
}
To show this, let us make an 
alternative derivation of the optimal matrix $\E^\l$.
Since both $\E^\l$ and $\C$ are symmetric, we can rewrite 
\eq{ClumsyCovEq} as
\beq{SmartCovEq}
\V_{\l\l'} = 2\tr[\C\E^{\l}\C\E^{\l'}],
\eeq
so we simply want to minimize $\tr \C\E^{\l}\C\E^{\l}$ subject to 
$\tr \P^\l\E^{\l}=1.$ Introducing a Lagrange multiplier $\lambda$, 
we wish to minimize the function
\beq{LagrangeEq}
\Ell = \tr\left[\C\E^{\l}\C\E^{\l} - 2\lambda(\P^\l\E^{\l}-1)\right].
\eeq
Requiring the derivatives with respect to the components of $\E^\l$ to vanish,
we obtain
\beq{ProofEq1}
\C\E^{\l}\C = \lambda\P^\l,
\eeq
so substituting the $\lambda$ that gives $\tr \P^\l\E^{\l}=1$ leaves us with the 
simple solution
\beq{SmartSolutionEq}
\E^\l = {1\over 2\F_{\l\l}}{\C^{-1}\P^\l\C^{-1}},
\eeq
where $\F$ is the Fisher information matrix given by \eq{FCeq}, {\ie}, 
$\F_{\l\l}=\tr[\C^{-1}\P^\l\C^{-1}\P^\l]/2$.
Comparing this with \eq{UselessSolutionEq}, we see that 
\eq{SmartSolutionEq} has the advantage that only a much smaller matrix
$(n\times n)$ needs to be inverted.

\subsection{...and lossless}
\label{LosslessSec}

We will discuss the properties of
this power spectrum estimation method at some length in 
Section~\ref{ExampleSec}, as well as apply it to the COBE data. 
Before doing this, however, we will now prove that this method is lossless 
in the sense defined above. (Our derivation of the method merely guaranteed that 
it was the best {\it quadratic} method, but did not rule out the possibility
that it destroys information and is inferior to some more nonlinear technique.)

The way we chose to normalize $\E_\l$ does clearly not affect the error bars with which
we can determine cosmological parameters, since multiplying 
the estimated power spectrum coefficients $\Ch_\l$ by some constants
(or indeed by any invertible matrix) will not change their information content
(see \cite{karhunen}). 
To simplify the calculation below, let us therefore scrap the 
$\F_{\l\l}$-factor in \eq{SmartSolutionEq} and define rescaled power
spectrum coefficients
\beq{yDefEq}
y_\l\equiv\x^t\E^\l\x,
\eeq
where
\beq{RescaledEeq}
\E^\l\equiv{1\over 2}{\C^{-1}\P^\l\C^{-1}}.
\eeq
Substituting \eq{RescaledEeq} into \eq{SmartCovEq}, we now
find that the covariance matrix for
$\y$ reduces to simply
\beq{yCovEq}
\expec{\y\y^t}-\expec{\y}\expec{\y}^t = \F,
\eeq
the Fisher matrix.
Arranging the true power spectrum coefficients $C_\l$ into a vector $\c$,
\eq{EstExpecEq} takes the simple form
\beq{EstExpecEq2}
\expec{\y} = \F\c.
\eeq
In other words, the window function matrix $\W$ of \eq{WindowDefEq} is also
equal to the ubiquitous Fisher matrix.
This means that if we use the vector 
\beq{ctDefEq}
\ct\equiv\F^{-1}\y
\eeq
to estimate the power spectrum $\c$, then this estimator will have the 
nice property that it is unbiased: 
\beq{UnbiasedEq2}
\expec{\ct}=\c.
\eeq
This Fisher-Cramer-Rao inequality (see \cite{karhunen} for a review) tells us that
the best an unbiased estimator can possibly do (in terms of giving small error
bars) is for its covariance matrix to equal 
$\F^{-1}$, the inverse of the Fisher matrix. 
Using \eq{yCovEq} and \eq{ctDefEq}, we find that 
this covariance matrix is precisely
\beq{OptProofEq}
\expec{(\ct-\c)(\ct-\c)^t} = \F^{-1}[\expec{\y\y^t}-\expec{\y}\expec{\y}^t]\F^{-1}=\F^{-1},
\eeq
so $\ct$ is indeed optimal in this sense.
In other words, we have found the best unbiased estimator of the power spectrum,
the one which gives the smallest error bars allowed by
the Fisher-Cramer-Rao inequality \footnote{
When a large fraction of the sky is covered, $\Ch_\l$
is effectively the average of $(2\l+1)$ independent multipoles,
so for $\l\gtrsim 50$, it will have a virtually perfect 
Gaussian distribution by the central limit theorem. 
This means that the post-compression
Fisher matrix is almost exactly $\F$, the value computed from the map, 
so that the power spectrum estimates retain all the small-scale
cosmological information. For the very lowest multipoles $\l\lesssim 4$, 
the Gaussian approximation becomes poor, so that it might be
desirable to supplement the power spectrum estimates with some 
linear measures of the large-scale power 
($a_{\l m}$-coefficients, say, as suggested in \cite{karhunen}) 
to ensure that every bit of information
is retained. This may be desirable anyway, to
avoid non-Gaussianity in the likelihood calculation, as will be
further discussed in Section~\ref{ParEstSec}.
}.
The fact that it turned out to be a simple quadratic
estimator is good news for CMB data analysis, since this means that it is much simpler
to implement in practice than for instance the highly nonlinear ML-method.

\section{A WORKED EXAMPLE: THE COBE DATA}
\label{ExampleSec}

In this section, we will discuss various aspects of how the method works. 
To prevent the discussion from becoming overly dry and abstract, we will
illustrate it with a worked example: application of the method 
to the 4 year COBE/DMR data.

\subsection{The COBE power spectrum}

We combine the 53 and 90 GHz channels (A and B) of the
COBE DMR 4 year data \cite{Bennett} into a single sky map
by the standard minimum-variance weighting, pixel by pixel.
We use the data set that was pixelized in galactic coordinates.
After excising the region near the galactic plane with
the ``custom cut" of the COBE/DMR team \cite{Bennett}, 
$n=3881$ pixels remain. As has become standard, we make no attempts
to subtract galactic contamination outside this cut.  We remove the
monopole and dipole and include this effect as described in the Appendix.

The resulting power spectrum is shown in \fig{ClFig} (top),
and is very similar to that extracted with 
the eigenmode method \cite{cobepow} --- we will discuss the 
relation between various
methods below. 
A brute force likelihood analysis of the 4 year data
set \cite{Hinshaw} gives a best fit normalization
of $\Qrmsps=18.4\,\mK$ for pure Sachs-Wolfe $n=1$ model,
corresponding to the
heavy horizontal line in the figure, and we used this as the fiducial
power spectrum when computing $\C$.
If this model were correct, we would expect approximately
$68\%$ of the data points to fall within the shaded
$1-\sigma$ error region.
As can be seen, the height of this region
(the size of the vertical error bars)
is dominated by cosmic variance
for low $\l$ and by noise for large $\l$.

\subsection{The window functions}

Our COBE example illustrates a number of general features 
regarding how the window function depends on the target multipole
and on the sky coverage, as shown in Figures 3 and 4. 
However, since we are conforming to the customary way of plotting 
window functions here, rather than to the definition of
\eq{WindowDefEq}, a few clarifying words regarding precisely what
is plotted are in order before proceeding.

\subsubsection{What they mean} 

As has become standard, the vertical axis in 
\fig{ClFig} shows not $\Cl$ but the relatively flat quantity 
\beq{DdefEq}
\D_\l\equiv \l(\l+1)\Cl/2\pi,
\eeq 
so we want to interpret the data points as weighted
averages of these quantities (rather than as weighted averages of
the $\Cl$-coefficients), with
the window function giving the weights.
In addition, we must take into account the fact that 
the COBE beam smearing suppresses the true multipoles $\Cl$ by the known factors 
$B_\l^2$ given by \cite{W94a}. We thus rewrite
\eq{CsumEq} as 
\beq{DsumEq}
\C = \expec{\x\x^t} = \N + \sum_\l \Pt^\l\D_\l,
\eeq
where we have defined $\Pt^\l\equiv\mu_\l\P^\l$ and 
\beq{muDefEq}
\mu_\l\equiv{B_\l^2\over\l(\l+1)/2\pi},
\eeq
and find the Fisher matrix for the $D_\l$-coefficients to be
\beq{FtDefEq}
\Ft_{\l\l'} \equiv\mu_\l\mu_{\l'}\F_{\l\l'}.
\eeq
Since each window function by definition must add up to unity, 
the correct normalization for the band-power estimators of $\D_\l$ 
is 
\beq{DhDefEq}
\Dh_\l\equiv\norm_\l \mu_\l y_\l = {\mu_\l\norm_\l\over 2}\x^t\C^{-1}\P_\l\C^{-1}\x,
\eeq 
where the normalization constants are defined by 
\beq{NormDefEq}
\norm_\l\equiv\left[\sum_{\l'}\Ft_{\l\l'}\right]^{-1}.
\eeq 
This implies that the mean is 
\beq{DexpecEq}
\expec{\D_\l} = \sum_{\l'}\Wt_{\l\l'}\D_{\l'},
\eeq
where the window function
\beq{WtdefEq} 
\Wt_{\l\l'}\equiv\norm_\l\Ft_{\l\l'}
\eeq
satisfies $\sum_{\l'}\Wt_{\l\l'}=1$,
and the covariance is 
\beq{DcovEq}
\expec{\D_\l\D_{\l'}}-\expec{\D_\l}\expec{\D_{\l'}} 
= \norm_\l\norm_{\l'}\Ft_{\l\l'}.
\eeq
For $\l=\l'$, this expression gives the (squared) 
error bars plotted on the data points $\Dh_\l$ in 
\fig{ClFig}. The horizontal point locations and the
corresponding horizontal bars give the means and 
{\rms} widths of the window functions $\Wt_{\l\l'}.$

\subsubsection{How they depend on the target multipole}

\Fig{LocationFig} shows the window functions for 
estimating the multipoles $\l_*=$10, 15 and 20.
As can be seen, their shape and width is  
more or less the same, so increasing the target multipole 
$\l_*$ merely translates them in $\l$-space.
This quantitative result is easy to understand in terms
of the quantum mechanics analogy made in \cite{window}: if the wave function
of a quantum particle on a sphere is required to vanish 
in certain regions, then its angular momentum distribution
(spherical harmonic coefficients) must have a certain minimum
width which is independent of the average angular momentum 
(in our case, independent of the target multipole $\l_*$).

\subsubsection{How they depend on the sky coverage}

The Heisenberg dispersion formula 
(``uncertainty relationship") tells us that this minimum
width is of order \cite{window}
\beq{RuleOfThumbEq}
\Delta\l \sim 1/\Delta\theta,
\eeq
where $\Delta\theta$ is the angular size in radians of the 
smallest dimension of the sky patch.
This simple scaling is quantitatively illustrated in
\fig{WidthFig}. For instance, comparing the two middle panels 
shows that adding a second hemisphere does not reduce the
width of the window (since $\Delta\theta$ remains unchanged), but
merely removes the power leakage from the even multipoles
(since the galaxy cut is approximately symmetric under reflection, 
and even and odd multipoles remain roughly orthogonal, since
they have opposite
parity).

\subsection{The essence of the method}

How does our method work?
In this section, we will see that it is quite straightforward to acquire
an intuitive understanding what the method does with the data
and why this improves the situation.
First we note that \eq{DhDefEq}
can be rewritten as 
\beq{DhEq2}
\Dh_\l\propto\z^t\P^\l\z,
\eeq
where the vector $\z$ is defined as
\beq{zDefEq}
\z\equiv\C^{-1}\x.
\eeq
Since $\z$ also consists of $n$ numbers, we can plot it as a sky map,
as is done in \fig{MapsFig}. Moreover, by the addition
theorem for spherical harmonics, we can factor the matrix $\P_\l$ as
\beq{PfactorEq}
\P^\l = \Y_\l\Y_\l^t,
\eeq
where the $n\times(2\l+1)$-dimensional matrix $\Y_\l$ is
defined as
\beq{YlDefEq}
\Y^\l_{im}\equiv Y_{\l m}(\rh_i).
\eeq 
Here and throughout, we let $Y_{\l m}$ denote
the real-valued spherical harmonics, which
are obtained from the standard spherical harmonics by replacing $e^{im\phi}$
by $\sqrt 2\sin m\phi$, $1$, $\sqrt 2\cos m\phi$ for $m<0$,
$m=0$, $m>0$
respectively.
Combining the last four equations, we find that
\beq{InterpEq1}
\Dh_\l\propto {1\over (2\l+1)}\sum_{m=-\l}^\l 
\left|\sum_i Y_{\l m}(\rh_i) z_i\right|^2,
\eeq
which we recognize as the method of 
expansion in truncated spherical harmonics
\cite{Hauser,W94b,W96a},
{\it but applied to the map} $\z$ {\it instead of the map} $\x$.

\Fig{MapsFig} compares the maps $\x$ and $\z$, and visual inspection 
reveals that although the small-scale features of $\x$ remain 
visible in $\z$, there are two obvious differences:
\begin{enumerate}
\item $\z$ looks high-pass filtered, with large-scale fluctuations
rendered almost invisible. 
\item The edges are softened by downweighting the pixels near
the galaxy cut, notably in the high signal-to-noise case.
\end{enumerate}
As described below, both of these phenomena have a simple intuitive 
explanation.

\subsubsection{High-pass filtering}

Since $\expec{\x\x^t}=\C$, it follows that
\beq{zCovEq}
\expec{\z\z^t} = \C^{-1}\expec{\x\x^t}\C^{-1}=\C^{-1}.
\eeq
In other words, those modes which had the most power in the
original map $\x$ have the least power in the map $\z$ and 
vice versa.
Computing $\Dh_\l$ from $\z$ with the simple spherical-harmonic
method is therefore akin to the standard way of
estimating a mean with inverse-variance weighting: first we re-weight
the numbers by dividing them by their variance, then we perform 
a straight average on the result. 
\Fig{TiltFig} illustrates the effect of this procedure in the Fourier
(multipole) domain, on the window functions.
The top panel shows the result of applying the 
straight spherical-harmonic method directly to $\x$.
The reason that the results are so poor is that 
the power spectrum $\C_\l$ falls rapidly with $\l$, so that
even though the ``red leak" from lower multipoles is small
on geometric grounds, the amount of large-scale power being aliased into 
the estimates of high multipoles is nonetheless comparable 
to the weak small-scale signal that we are trying to measure.
The middle panel shows the optimal method, {\ie}, applying the
straight spherical-harmonic method to $\z$.
Since the power spectrum of $\z$ is ``tilted" to suppress
the large-scale power, the troublesome red leak is seen to be virtually
eliminated: it no longer matters that some fraction of the large-scale 
power in $\z$ gets aliased down to small scales, since there is 
is so little large-scale power there in the first place.

The bottom panel shows the window function that would result if the
signal-to-noise ratio was about 500 times higher, roughly
corresponding to what is expected for the upcoming {\it Planck} satellite.
Here the high-pass filtering is seen to be more extreme, and the window
function is seen to be slightly narrower still. 
If the pixel noise is uniform and uncorrelated, then 
$\C$ will clearly become proportional to the identity matrix if
we let the signal-to-noise ratio approach zero. This means that if the
noise in a map is orders of magnitude greater than the signal, then 
$\z\propto\x$ and the best pixel weighting becomes
to do nothing at all, leaving the map as it is. 
In this sense, the three maps in \fig{MapsFig}
form a progression of $\z$-maps corresponding to increasing 
signal-to-noise.

\subsubsection{Edge tapering}

The second salient feature of the method, ``feathering" the edges,
is also easy to understand in terms of the quantum mechanics 
analogy given in \cite{window}. For the window function to be narrow,
we loosely speaking want the pixel weighting 
to be narrow in the Fourier (multipole) domain, 
so to avoid excessive ``ringing" in Fourier space, 
the weighting in real space should be continuous and smooth.
This standard signal-processing procedure is also known as
apodizing, and is routinely used in the analysis of one-dimensional
time series data \cite{Press}.
An alternative way to see why the edges are softened is to consider
the noise map, {\ie}, the map of the standard deviations of the pixels
in $\z$. This is simply the square root of the diagonal part of its
covariance matrix, {\ie}, $\Delta z_i = (\C^{-1})_{ii}^{1/2}$.
As is readily verified even for a simple one-dimensional array of 
pixels, the diagonal elements of $\C^{-1}$
will be smaller near the edges even though all diagonal 
elements of $\C$ itself are the same (since $\C_{ij}=c(\rh_i\cdot\rh_j)$
for some correlation function $c$ when the noise is uniform).

Just as with the high-pass filtering, the degree of edge tapering is seen to 
increase with the signal-to-noise ratio $S/N$.
As mentioned, $S/N\to 0$ gives $\C\propto\I$, {\ie}, no edge softening
at all, whereas the noise map tends to zero at the edges
when $S/N\to\infty$. In this sense, our new method strikes a balance
between the all-out apodization of the eigenmode method \cite{window}
and the
{\it laissez-faire} approach of the Hauser-Peebles method \cite{Hauser}, 
with the amount of softening depending on what is affordable given the noise.
Clearly, if $S/N\sim 0$, the sole source of variance in
$\Dh_\l$ is the noise, so the widths of the window functions
(and hence the cosmic variance leakage from unwanted aliased multipoles)
is irrelevant,
and we simply wish to weight all pixels equally (or by the inverse of 
their noise variance if the noise is non-uniform).
If there is very little noise, on the other hand, we can go to great lengths
to narrow down the window functions. Indeed, the excellent
S/N of {\it Planck} drives the algorithm to ``deconvolve" the power spectrum 
to obtain an even narrower window in the bottom panel of \fig{MapsFig}
than in the middle one. 

\subsection{Speed issues} 

A quick glance at \eq{DhDefEq} might give the impression that
inversion and multiplication of $n\times n$ matrices are required
to compute the estimated power spectrum $\Dh_\l$, which are devastatingly
slow $n^3$ operations. Fortunately, this impression is
deceptive, since \eq{DhEq2} shows that merely 
a matrix-vector product (ordo $n^2$), a vector-vector product
(ordo $n$) and a linear equation solution stemming from \eq{zDefEq} 
need be carried out to obtain the raw power estimates. 
The latter is also an ordo $n^2$ operation when using 
an iterative approach, {\eg}, the Gauss-Seidel method.
Moreover, there is no need to store the $n\times n$ elements
of $\P^\l$ or $\C$, since they can be rapidly computed on the fly, as needed, 
for instance with cubic spline interpolation, so the storage
requirements here are merely ordo $n$.

Similar manipulations enable great time savings when computing the
Fisher matrix, which as we saw gave both the window functions and
the covariance matrix of the power estimates by simple rescalings
of its rows and columns. 
Suppose we are interested in the power spectrum up to some 
monopole $\lmax$. There are $(\lmax+1)^2$ spherical harmonics
with $\l\leq\lmax$, so let us define the 
$n\times(\lmax+1)^2$-dimensional matrix $\Y$ just as in 
\cite{brute}, as
\beq{YdefEq}
\Y_{i\lambda}\equiv Y_{\l m}(\rh_i).
\eeq
Here we have combined $\l$ and $m$ into a single index
$\lambda\equiv\l^2+\l+m+1 = 1, 2, 3, ...$
Using \eq{FCeq}, \eq{PfactorEq} and the fact that a trace of a product of matrices
is invariant under cyclic permutations, we obtain the useful result that
\beq{FastFisherEq}
\F = {1\over 2}\sum_{m=-\l}^\l\sum_{m'=-\l'}^{\l'} (\V_{\lambda\lambda'})^2,
\eeq
where the matrix $\V$ is defined as
\beq{VdefEq}
\V\equiv\Y^t\C^{-1}\Y.
\eeq
Since $\C^{-1}\Y$ can be solved iteratively for each 
spherical harmonic (row of $\Y$) separately, and since the structure of 
\eq{FastFisherEq} shows that there is no need to ever load the entire 
$\V$-matrix into memory all at once, the computation of $\F$ thus 
poses no significant computer memory challenges and lends itself well
to parallelization. The computation of the the noise bias corrections
$b_\l$ are straightforward to accelerate in an analogous way.
Note that \eq{FastFisherEq} also proves that all elements of the
Fisher matrix are non-negative, which among other things means that
the optimal method will never give window functions that go negative.

Alternatively, if an approximation of $\F$ is deemed satisfactory, it
can of course be estimated quite rapidly by computing 
$\Dh_\l$ from a large number of Monte-Carlo skies.

Finally, it should be noted that the calculation of the parameter
covariance matrix takes roughly the same amount of time for this method as 
it does for the Hauser-Peebles method. For that case, 
the covariance matrix for the power spectrum estimates can be rewritten in
the form of the right hand side of \eq{FastFisherEq}
but with $\V\equiv\Y^t\C\Y$ instead of $\Y^t\C^{-1}\Y$
Since matrix multiplication takes roughly as long as matrix inversion, 
this shows that although the simple approach of 
estimating the power spectrum with a straight truncated spherical harmonic 
expansion of the original map is inferior in terms of 
destroying information, it is not substantially faster.

\section{FURTHER DOWN THE PIPELINE: WHAT TO DO WITH $\Ch_\l$}
\label{OutlookSec}

In this section, we will briefly discuss that part of the data analysis pipeline
in Figure 1 which lies beneath the power spectrum estimation 
step. As discussed in the introduction, there are many reasons to 
plot the power spectrum for direct visual inspection. The power spectrum estimates
also constitutes a small and manageable data set that retains all the cosmological 
information from the original map in Gaussian models, and is therefore useful
as a starting point when constraining cosmological models and model parameters.
We will now discuss these two applications in turn.

\subsection{Using $\Ch$ for ``chi-by-eye"}

Given the vector of raw power spectrum estimates $\yt$, 
where $\tilde{y}_\l\equiv\mu_\l y_\l$, $\l=2,...,\lmax$, and $\y_\l$ is defined
as in Section~\ref{LosslessSec},
there are a number of ways to take linear combinations of them,
normalize them 
and plot them with vertical and horizontal error bars as in \fig{ClFig}. 
We will now discuss some natural ones briefly and comment on their relative merits.

\subsubsection{Raw estimates}

The minimalistic approach is of course to do nothing
and simply plot $\yt$ as is. This is 
the simple approach taken in the top panel of \fig{ClFig}. 

\subsubsection{Band power}

One disadvantage of the previous approach is the profusion of data points,
sampling the power spectrum with a much narrower $\l$-spacing 
(in intervals $\Delta\l=1$) than the scale on which typical 
models are expected to vary noticeably
(typically $\Delta\l\gg 10$, at least for $\l\gg 50$).
A simple remedy is of course to average the raw power estimates
in appropriate bands, essentially smoothing  
\fig{ClFig}.

\subsubsection{Deconvolved power}

The exact opposite approach is to ``un-smooth" or deconvolve the
power spectrum so that all data points become uncorrelated and 
all window functions become Kronecker delta functions.
Although this can be formally accomplished by computing 
$\Ft^{-1}\yt$ as in Section~\ref{LosslessSec}, where we did this
to prove that our method had retained all the information 
that there was, this 
is of course a terrible idea in practice, since 
incomplete sky coverage makes the Fisher matrix nearly 
singular. The result would be a plot with gigantic error bars,
which is simply Nature's way of telling us that we cannot really 
measure the power spectrum with a resolution below the natural 
scale set by $\Delta\l\sim 1/\Delta\theta$.

\subsubsection{Uncorrelated data points}

A more fruitful way of producing uncorrelated data points is 
to plot the numbers in the vector $\X^t\yt$, where the rows of the matrix 
$\X$ are the solution vectors to the generalized eigenvalue problem
\beq{GenEigenEq}
\U\X=\Ft\X\Lambda
\eeq
for some symmetric matrix $\U$. Here $\Lambda$ is a diagonal matrix 
containing the eigenvalues.
It is easy to show that for any choice of $\U$, all the new data points
will be uncorrelated with unit variance (they are of course appropriately
rescaled when plotted), and the new window function 
matrix will be simply $\X^t$.

\subsubsection{Principal components}

One special case of the above approach is to choose $\U=\I$, the identity
matrix, which reduces it to a so-called principal component
analysis. By sorting the new data points (the principal components) 
by their eigenvalues, one can rank them from best
to worst and throw away a substantial number of essentially redundant ones,
thereby getting around the problem that there are, loosely speaking, 
too many data points for them to all be uncorrelated and yet not in some
sense pathological.

\subsubsection{Hamilton coefficients}

\label{UncorrSec}

In two recent papers \cite{H97a,H97b}, an extensive study of 
choices of $\U$ was carried out for the related problem of 
how to present the power spectrum measured from galaxy 
surveys. It was found that most choices, including 
that of principal component analysis, are unfortunately 
{\it not} particularly useful, since they tend to produce 
quantities whose window functions are ``unphysical" in the sense
of being extremely broad and often negative.
However, it was found that a certain limiting case
\cite{H97b} tends to produce 
nice and clean window functions, and in addition eliminates the
need to solve an eigenvalue problem. These cases correspond
to factoring the Fisher matrix as
\beq{CholeskyEq}
\Ft = \M\M^t
\eeq
for some matrix $\M$, and chosing $\X=\M^{-1}$.
Of the infinitely many choices of $\M$, three are
particularly attractive \cite{texas96}:
\begin{enumerate}
\item
If one requires $\M$ to be {\it lower-triangular}, 
in which case $\F=\M\M^t$ corresponds to a Cholesky decomposition,
the COBE case gives the narrow and non-negative window functions
in the middle panel of \fig{UncorrWindowFig}, with side lobes only to the right.
\item
Similarly, one could obtain window functions with
side lobes only to the left by chosing $\M$ {\it upper-triangular}.
\item
A third choice, which is the personal favorite of the author,
is choosing $\M$ {\it symmetric}, which we write as $\M=\F^{1/2}$.
The square root of the Fisher matrix is seen to give 
beautifully symmetric window functions
(\fig{UncorrWindowFig}, bottom) that are not only non-negative, but
also even narrower than the original (top), which has roughly the 
bottom profile convolved with itself.
\end{enumerate}
The bottom panel of
\Fig{ClFig} shows the COBE power spectrum plotted with
this last method.
These 29 data points thus contain all the cosmological information
from COBE, distilled into 29 chunks that are not only 
collectively exhaustive (jointly 
retaining all the cosmological information),
but mutually exclusive (uncorrelated) as well.
The above-mentioned band averaging now has the nice feature that
the band powers will automatically be uncorrelated as well,
so to reduce scatter, the 29 measurements 
have been binned into 8 bands in Table 1
and \fig{ExperimentsFig}.
The data and references for the other experiments 
plotted can be found in recent compilations~\cite{Lineweaver,Rocha}.

\subsubsection{Negative power?}

We close this discussion by remarking that with all these
approaches, it is possible for data points to be negative,
which may annoy certain readers since the true power spectrum is
by definition nonnegative. 
It should be emphasized that this is a purely stylistic 
issue of no scientific importance whatsoever. As we proved in 
Section~\ref{LosslessSec}, the raw power estimates contain 
all the cosmological information there is, regardless of whether
we plot them or not. The total power that we measure in a given 
multipole will always be nonnegative, and the reason that
negative values can occur in figures is simply that we are plotting 
the difference between two positive quantities: what we measure
and the noise bias. Plotting the the sum of $y_\l$ and the noise bias would 
result in figures guaranteed to be free of negative points, 
clearly containing exactly the same information as those described
above since the noise bias is a set of known constants, 
but having the undesirable property of being biased upward.
Alternatively, some non-linear mapping could be used to guarantee 
positivity of the plotted points, but for subsequent analysis
as outlined in the following section, we obviously want to retain
our simple quadratic estimators to avoid complicating the statistical
properties of the measured power spectrum.

\subsection{Using $\Ch_\l$ for parameter estimation}

\label{ParEstSec}
The second, and arguably most important, use for the
power spectrum estimates is to make it possible 
to place sharp quantitative constraints on cosmological models
and their parameters.

\subsection{The simple-minded approach: maximum likelihood}

As described in \cite{karhunen}, likelihood analysis has emerged as one
of the most popular data analysis tools in cosmology because it 
is often simple to implement and in addition is the best method
in certain asymptotic situations.
In our case, measuring say the 11 CDM-parameters of 
\cite{JKKS2} via a direct likelihood analysis
using our power spectrum estimates $\Dh_\l$ would unfortunately
be extremely cumbersome numerically for the case of a megapixel 
CMB map. The reason is that it would, even in the crude and poor approximation
that $\Dh_\l$ is Gaussian all the way down to the lowest multipoles,
require computing their $\lmax\times\lmax$ covariance matrix 
at a grid of points in parameter space. Although this covariance matrix
is relatively small, we saw above that its computation was so time-consuming 
that it would be preferable to compute it only once (or a few times).
In addition, the central limit theorem only guarantees that
the multipole estimates have Gaussian probability distributions
for $\l\gtrsim 50$, so the exact likelihood function 
will be extremely cumbersome to compute when the lower
multipoles are included in the analysis.

\subsection{Why chi-squared can be just as good}

This situation is similar to that in Section~\ref{FirstGuessSec}, 
where we found that the ML-method
was numerically undesirable and hoped that a simpler method 
would provide error bars that were equally good or better.
In that instance, a simple quadratic method came to the rescue, 
and we will argue that the story repeats itself here in the
final step in the pipeline: that the simple 
(quadratic) chi-squared method 
is likely to be, if not better, at least almost as good as the
ML-method when fitting parameters to the power spectra measured from
megapixel CMB maps.

The Karhunen-Lo\`eve data compression 
problem has been generalized and solved  
for the case at hand here \cite{karhunen}:
we have a data vector (say, the raw power spectrum estimates $\y$)
whose mean 
\beq{yMeanEq}
\m\equiv\expec{\y}
\eeq
and covariance matrix 
\beq{yCovEq2}
\M\equiv \expec{\y\y^t}-\expec{\y}\expec{\y^t}
\eeq
both depend in a known way on the parameters that we wish to estimate
(say $h$, $\Omega$, {\etc}).
When changing the parameters, the likelihood function of course 
changes for two reasons:
\begin{itemize}
\item Because $\m$ changes
\item Because $\M$ changes 
\end{itemize}
Another way of phrasing this is that the information that the ML-method
extracts from the parameters comes from two sources: the parameter dependence
of $\m$ and that of $\M$. 
In \cite{karhunen}, it was found that for our case, as long as the 
parameters were well constrained (as is anticipated
for both {\it MAP} and {\it Planck} \cite{JKKS2}),
{\it virtually all the information came from $\m$, not from $\M$.}
This means that, in the Gaussian approximation, 
the likelihood curve near the maximum is well approximated by
\beqa{Chi2Eq}
\nonumber
L&\simpropto&[\det\M]^{-1/2}\exp\left[-{1\over 2}(\y-\m)^t\M^{-1}(\y-\m)\right]\\
 &\simpropto&\exp\left[-{1\over 2}(\y-\m)^t\M^{-1}(\y-\m)\right],
\eeqa
where $\M$ is a merely a {\it constant} matrix evaluated somewhere near
the best fit point in parameter space.
Maximizing this is of course amounts to making a simple
chi-squared model fit to the plotted power spectrum, and requires 
no overly tedious numerical calculations at all: the mean 
vector predicted by a given model is simply given 
by \eq{EstExpecEq2}, {\eg}, by the Fisher matrix (which is computed once and for all)
times the model power spectrum.
If the best fit power spectrum turns out to be so different from
the one assumed when computing $\F$ that the accuracy
of \eq{Chi2Eq} comes into doubt, the best fit parameter values can of course
be used to repeat the entire procedure iteratively.
    Needless to say, the approach outlined in this section needs to be 
implemented and tested in considerable detail before any strong statements
can be made about its feasibility. For instance, there is no guarantee
that the chi-squared method (nor, for that matter, the ML-method) will 
give unbiased parameter estimates, so Monte-Carlo calibrations are
necessary in either case. The fact that the multipole estimates with $\l\ll 50$ are
non-Gaussian also means that chi-squared parameter estimates
made with the Gaussian expression of \eq{Chi2Eq} are likely to give error bars
on the parameters that are slightly above the theoretical minimum.
We merely conclude this section by saying that it appears {\it plausible} that 
a simple chi-squared approach in the final step of Figure 1 
will give error bars on cosmological parameters
that are almost as small as theoretically possible.


\section{CONCLUSIONS}
\label{ConclusionsSec}

We have presented a new method for power spectrum estimation from 
CMB maps and argued that it is the best choice for the box labeled 
$\E$ in Figure 1:
\begin{itemize}
\item Just as the mapmaking step above it, it can compress the data set by a large factor 
while retaining all the cosmological information.
\item It is simple enough to be computationally feasible in practice even for 
future megapixel sky maps.
\item The statistical properties of the power spectrum estimates are straightforward to 
compute, and using a simple chi-squared parameter fitting approach 
in the bottom box in the figure is likely to give error bars on the parameters that 
are almost as small as theoretically possible.
\end{itemize}
We illustrated the method details by applying it to the 4 year COBE/DMR data.
It roughly speaking involves subjecting the map to a high-pass filter 
and some edge softening, and then analyzing the resulting map with 
the Hauser-Peebles method, {\ie}, with a straight
expansion in truncated spherical harmonics.
It reduces to the Hauser-Peebles method in
the limit of zero signal-to-noise. Both of these methods 
require approximately the same amount of CPU time.
When the signal-to-noise is very 
high, it becomes quite similar to the maximum spectral resolution method
\cite{window}. 
However, even in this limit, it has the advantage (in addition to being
strictly lossless, which the method presented 
in \cite{window} is not) that it does not require the solution
of an eigenvalue problem, thus being much faster. 
   
It has recently been shown that there is a mapmaking method
that is both lossless \cite{mapmaking} and computationally feasible 
\cite{W96b}.
Combining that with the present results, we conclude
that all aspects of the ``main tube" of the data analysis pipeline
are now under control, making it plausible that future CMB missions 
can deliver the promise of accurate measurements of cosmological parameters
not merely in principle but also in practice, without floundering on computational 
difficulties, if all other aspects of the problem are as simple as possible.
By this last caveat, we mean the following rosy scenario:
\begin{itemize}
\item There are no unforeseen systematic errors.
\item Non-Gaussian and anisotropic foregrounds can be removed down
to a tolerable level already in the map stage.
\item The CMB fluctuations are Gaussian.
\item The true model will turn out be be something similar to 
what we expect, so that the power spectrum really 
contains information about our classical cosmological parameters.
\end{itemize}
All of these issues require substantial amounts on work before 
we can trust our fledgling data analysis pipeline
in Figure 1 to be able to make the most of the real CMB
data sets that await us.

\bigskip
The author wishers to thank Andrew Hamilton for useful discussions.
Support for this work was provided by
NASA through a Hubble Fellowship,
{\#}HF-01084.01-96A, awarded by the Space Telescope Science
Institute, which is operated by AURA, Inc. under NASA
contract NAS5-26555.
The COBE data sets were developed by the NASA
Goddard Space Flight Center under the guidance of the COBE Science Working
Group and were provided by the NSSDC.


\section*{APPENDIX: MONOPOLE AND DIPOLE REMOVAL}

Since the monopole and the kinematic dipole of the CMB 
are orders of magnitude
larger than the multipoles of cosmological interest 
$(\l\geq 2)$, one customarily removes them from the map
$\x$ prior to any subsequent analysis. 
Indeed, the monopole cannot even be measured by 
differential experiments (such as COBE/DMR).

\subsection{The problem}

Suppose that all multipoles $\l<\l_0$ are removed
from the data set. This means that the data set 
at our disposition, say $\xt$, is given by
\beq{xtDefEq}
\xt\equiv\PP\x, 
\eeq
where $\PP$ is a projection matrix that projects onto the
subspace orthogonal to these multipoles.
If the columns of a matrix $\Z$ form an orthonormal 
($\Z^t\Z=\I$) basis for the space of
these unwanted multipoles\footnote{Such a matrix $\Z$ is readily constructed by 
starting with the unwanted columns in the multipole matrix $\Y$
(the first 4 columns if $\l_0=2$)
and orthonormalizing them with 
Gram-Schmidt or Cholesky procedure \cite{brute}.}, then
\beq{PiDefEq}
\PP=\I-\Z\Z^t.
\eeq
The covariance matrix of the available data is therefore
\beq{CtDefEq}
\Ct=\expec{\xt\xt^t}=\PP\C\PP^t.
\eeq
Since $\Z$ has $\l_0^2$ columns, $\PP$ and $\Ct$ have 
rank $n'\equiv n-\l_0^2$.
Hence $\Ct$ is singular and non-invertible, and the
method we have presented cannot be applied in its most 
straightforward implementation.
This problem occurs simply because the numbers in $\xt$ are not 
independent, \ie, since $\l_0^2$ of them can be expressed as linear 
combinations of the others.

\subsection{Three solutions}

There are three different ways of incorporating this complication
into our analysis:
\begin{enumerate}
\item Simply throw away $\l_0^2$ of the data points in $\xt$.
All the remaining data points are independent, so the 
resulting covariance matrix ($\Ct$ with the corresponding rows and
columns removed) is invertible. The drawback of this approach is
that the covariance matrix tends to be poorly conditioned, 
which may give numerical difficulties if $n$ is very large.
\item Use $\C$ rather than $\Ct$ for the calculations, 
but choose a fiducial power spectrum where $C_\l$ is very large
for $\l<\l_0$. Then the method itself will remove the unwanted multipoles
with great accuracy. This approach is of course not strictly 
correct, but is easy to implement and an excellent approximation
in most realistic cases.
\item Follow our method to the letter, 
but in place of $\Ct^{-1}$ (which is undefined), 
use the ``pseudo-inverse" of $\Ct$, the matrix defined as
\beq{MdefEq}
\M\equiv \PP\left[\Ct+\eta\Z\Z^t\right]^{-1}\PP^t.
\eeq
\end{enumerate}

\subsection{The pseudo-inverse approach}

We will now show that the definition of $\M$ is independent 
of the choice of the constant $\eta\ne 0$ 
and that this pseudo-inverse approach 
gives the desired optimal results.
Let $\R$ be an orthonormal $n\times n$ 
matrix whose first $\l_0^2$ columns equal the columns of $\Z$.
In other words, the columns of $\R$ form an orthonormal basis
and split into two blocks, corresponding to the   
unwanted and wanted multipoles, respectively.
In this new basis, the projection matrix $\PP$ takes the simple form
\beq{RPReq}
\R^t\PP\R=
\left(
\begin{tabular}{cc}
\0&\0\\
\0&\I
\end{tabular}
\right),
\eeq
where the upper left block in this $n\times n$ matrix 
has size $\l_0^2\times\l_0^2$, {\etc} In the same basis, 
$\Ct$ therefore takes the form
\beq{RCReq}
\R^t\Ct\R = (\R^t\PP\R) (\R^t\C\R) (\R^t\PP\R)
=
\left(
\begin{tabular}{cc}
\0&\0\\
\0&$\Ctt$
\end{tabular}
\right)
\eeq
for some non-singular $n'\times n'$ matrix $\Ctt$.
It thus follows that 
\beqa{RMReq}
\R^t\M\R&=& 
(\R^t\PP\R)\left[\R^t\Ct\R+\eta(\R^t\Z\Z^t\R)\right]^{-1}\R^t\PP^t\R\nonumber \\
&=&
\left(\begin{tabular}{cc}\0&\0\\ \0&\I\end{tabular}\right)
\left[
\left(\begin{tabular}{cc}\0&\0\\ \0&$\Ctt$\end{tabular}\right)
+\eta
\left(\begin{tabular}{cc}\I&\0\\ \0&\0\end{tabular}\right)
\right]^{-1}
\left(\begin{tabular}{cc}\0&\0\\ \0&\I\end{tabular}\right)
\nonumber\\
&=&
\left(\begin{tabular}{cc}\0&\0\\ \0&\I\end{tabular}\right)
\left(\begin{tabular}{cc}$\eta^{-1}$\I&\0\\ \0&$\Ctt^{-1}$\end{tabular}\right)
\left(\begin{tabular}{cc}\0&\0\\ \0&\I\end{tabular}\right)
= 
\left(\begin{tabular}{cc}\0&\0\\ \0&$\Ctt^{-1}$\end{tabular}\right),
\eeqa
independent of $\eta$. 
Combining the last three equations, we see that 
$\M\Ct=\Ct\M=\PP$, which implies that 
$\PP[\M\Ct-\I]\PP^t=\0$. Thus although there cannot be a matrix $\M$
such that  $\M\Ct-\I=\0$, our choice of $\M$ comes close to being an
inverse in the sense that the equation $\M\Ct-\I=\0$
becomes true when you project out the unwanted multipoles.

Let us now show that the pseudo-inverse approach is lossless.
Since
\beq{aDefEq}
\R^t\xt = (\R^t\PP\R)(\R^t\x) = 
\left(\begin{tabular}{c}\0\\ \a\end{tabular}\right)
\eeq
for some $n'$-dimensional vector $\a$, we can compute the
Fisher matrix directly 
from $\a$, since it clearly contains all the information.
Its covariance matrix is $\expec{\a\a^t}=\Ctt$, 
whose derivatives are given by
\beq{CttDerivEq}
\left(\begin{tabular}{cc}\0&\0\\ \0&${\partial\Ctt\over\partial C_\l}$\end{tabular}\right)
=\R^t{\partial\Ct\over\partial C_\l}\R
=\R^t\PP\P^\l\PP\R.
\eeq
Since $\PP\P^\l=\0$ for $\l<\l_0$, we of course have no information
about these multipoles.
The power spectrum Fisher matrix is given by
\beqa{CorrectedFisherEq}
\F_{\l\l'} &=& {1\over 2}\tr\left[
\Ctt^{-1}{\partial\Ctt\over\partial C_\l}
\Ctt^{-1}{\partial\Ctt\over\partial C_{\l'}}\right]
\nonumber\\
&=&{1\over 2}\tr\left[
(\R^t\M\R)(\R^t\PP\P^\l\PP\R)
(\R^t\M\R)(\R^t\PP\P^{\l'}\PP\R)\right]
\nonumber\\
&=&{1\over 2}\tr\left[\M\P^\l\M\P^{\l'}\right],
\eeqa
where in the last step, we used that $\R\R^t=\I$ and $\M\PP=\PP\M=\M$.
Using $\M\Ct=\PP$,  the covariance matrix of \eq{SmartCovEq}
reduces to
\beqa{CorrectedCovarEq}
\V_{\l\l'} &=&
{1\over 2}\tr\left[\Ct\M\P^\l\M\Ct\M\P^{\l'}\M\right]
\nonumber\\
&=&{1\over 2}\tr\left[\M\P^\l\M\P^{\l'}\right]=\F_{\l\l'},
\eeqa
so the argument of Section~\ref{LosslessSec} shows
that the smallest possible error bars are attained.

\subsection{The pseudo-inverse in practice}

Since $\Z$ contains merely a few columns, computing $\M$ using \eq{MdefEq} 
would be virtually as fast as computing $\C^{-1}$, and 
neither Cholesky decomposition nor iterative methods 
for computing the map $\z=\M\x$ suffer any noticeable speed loss
because of the monopole/dipole complication.
For instance, for the simple case $\l_0=1$, the pseudo-inverse is computed by
simply
\begin{enumerate}
\item subtracting the mean from all rows and columns of the matrix,
\item adding the constant $\eta$ to all the matrix elements,
\item inverting the matrix, and
\item again subtracting the mean from all rows and columns.
\end{enumerate}
Note that $\eta$ need not be small. As the result is independent of
$\eta$, it is numerically advantageous to make the monopole 
comparable to other eigenvalues, which corresponds to chosing 
$\eta=a/n$, where $a$ is the order of magnitude of a typical 
matrix element.\footnote{
This important special case $\l_0=1$ also applies to the related problem
of making CMB maps (symbolized by the box marked $\W$ in
Figure 1) from differential measurents, where one needs
to ``invert" a matrix which is singular because its 
rows and columns have zero mean. 
It is easy to show that for the COBE mapmaking method, 
the lossless map \cite{mapmaking} is obtained by using the pseudo-inverse
described here. 
When making their maps \cite{Bennett}, the COBE/DMR team
regularized the inversion by adding $\eta\I$ to their matrix, 
choosing $\eta$ very small. 
A better way of is therefore to add 
$\eta$ to {\it all} the $n\times n$ matrix
elements, not merely to the diagonal, and to choose 
$\eta=a/n$ rather than infinitesimal.
}


\begin{table}
\begin{tabular}{ccccccc}
Band&$\l_*$&$\expec{\l}$&$\Delta\l$&$\delta T$&$-1\sigma$&$+1\sigma$\\
\hline
1&2&2.1&0.5& 8.5& 0&24.5\\
2&3&3.1&0.6&28.0&17.7&35.5\\
3&4&4.1&0.7&34.0&26.8&40.0\\
4&5-6&5.6&0.9&25.1&18.5&30.4\\
5&7-9&8.0&1.3&29.4&25.3&33.0\\
6&10-12&10.9&1.3&27.7&23.2&31.6\\
7&13-16&14.3&2.5&26.1&20.9&30.5\\
8&17-30&19.4&2.8&33.0&27.6&37.6\\
\end{tabular}
\caption{
The COBE/DMR power spectrum 
$\delta T\equiv [\l(\l+1)C_\l/2\pi]^{1/2}$ in $\mu K$.}
\label{Table1}
\end{table}


\onecolumn

 
\clearpage
\begin{figure}[phbt]
\centerline{{\vbox{\epsfxsize=14cm\epsfbox{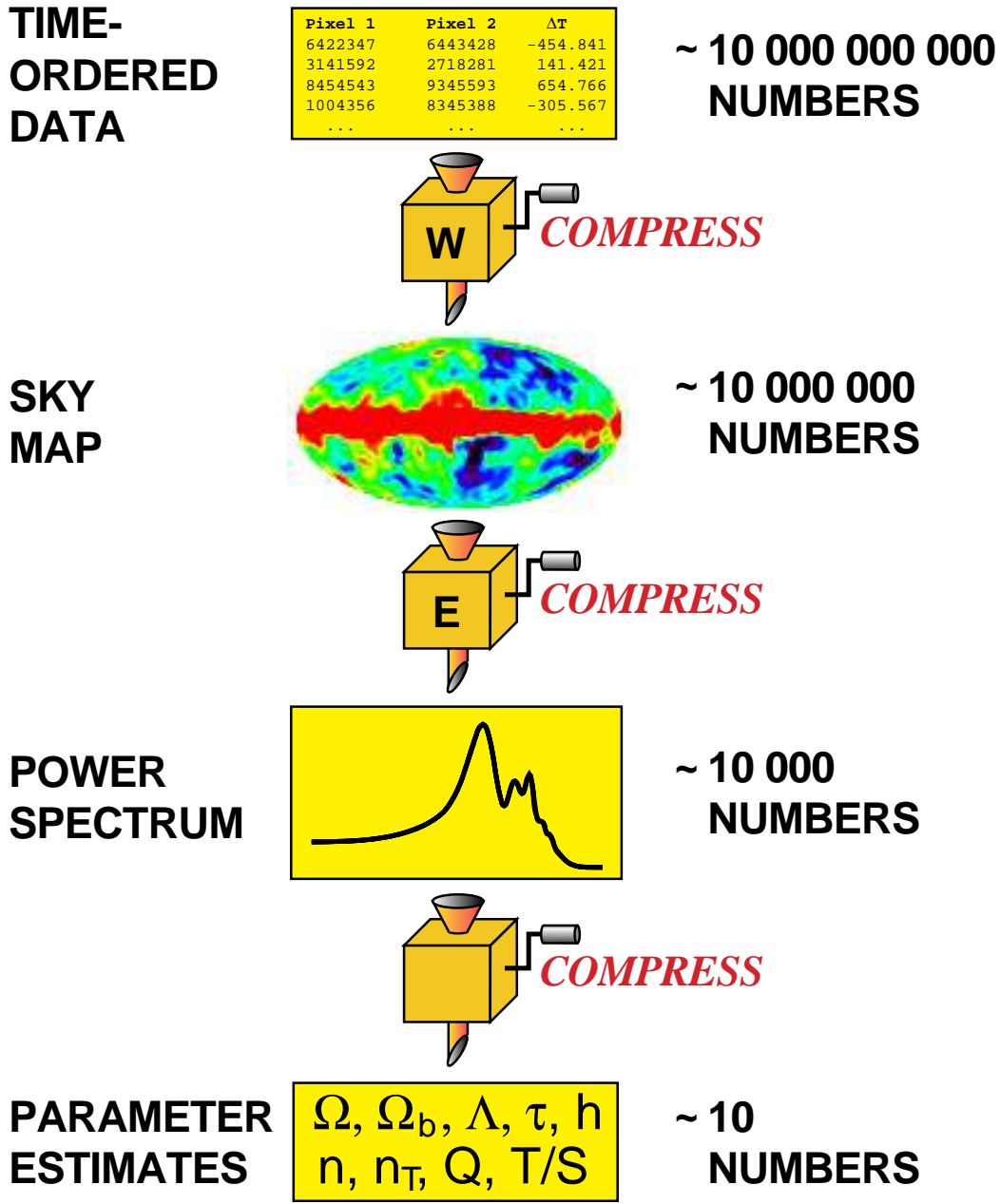}}}}
\vskip1.0cm
\label{PipelineFig}
\caption{
Power spectrum estimation as an intermediate data-compression 
step in converting the raw data from a future megapixel experiment
to measurements of cosmological parameters. 
If all three data compression steps are lossless, 
then this data analysis pipeline will measure the parameters with
just as small error bars as a (computationally unfeasible) likelihood 
analysis measuring the parameters directly from the time-ordered data.
}
\end{figure}

\clearpage
\begin{figure}[phbt]
\centerline{{\vbox{\epsfxsize=19cm\epsfbox{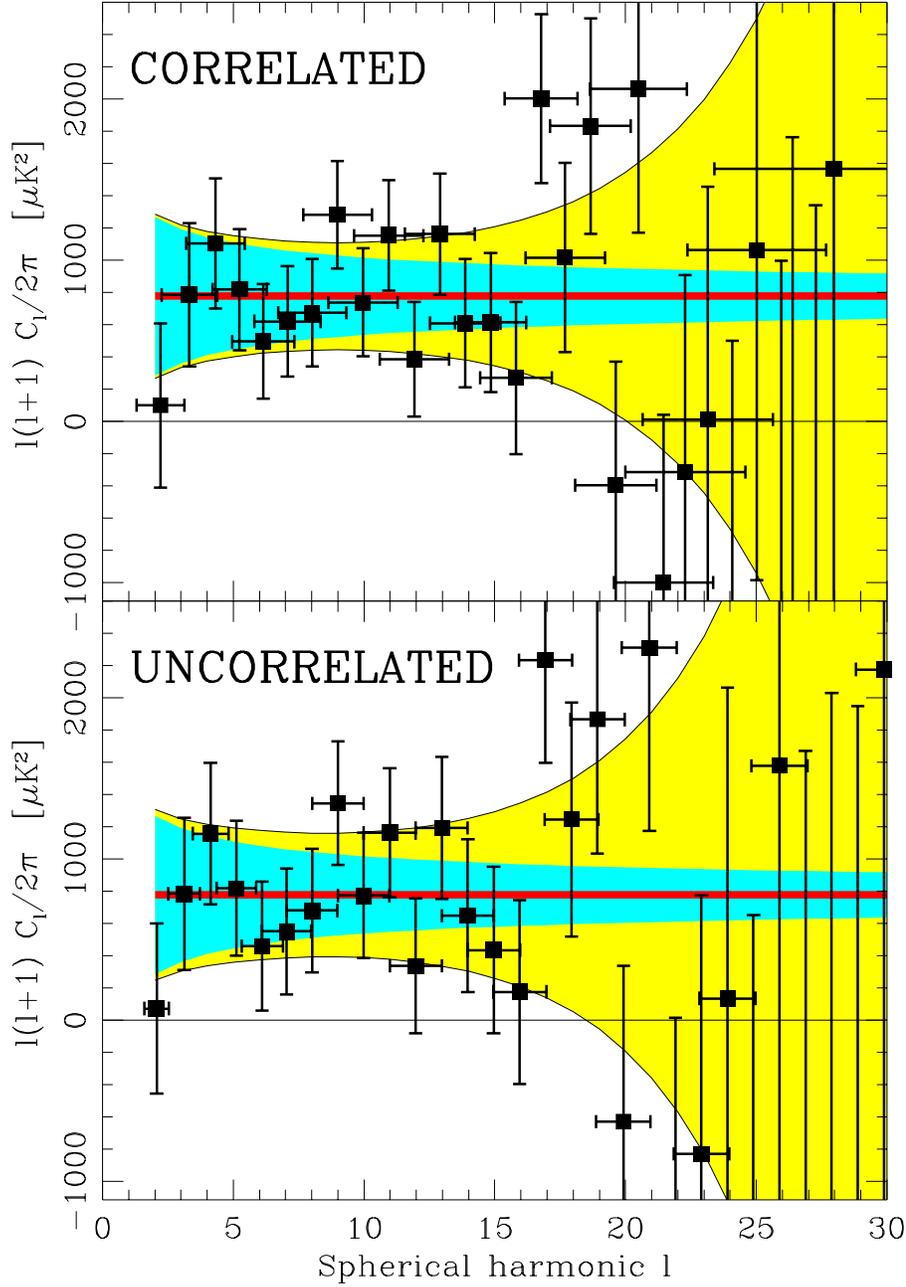}}}}
\caption{
The power spectrum estimated from the COBE/DMR 4 year data.
}
The observed multipoles $D_\l = \l(\l+1)C_\l$ are plotted with
$1-\sigma$ error bars using our basic method (top)
and made uncorrelated with the $\F^{1/2}$-method 
of Section~\ref{UncorrSec} (bottom).
The vertical error bars include both
pixel noise and cosmic variance, and the horizontal bars
show the width of the window functions used.
If the true power spectrum is given by
$n=1$ and $Q_{rms,ps}=18.4\mK$ (the heavy horizontal line),
then the shaded region gives the $1-\sigma$ error bars and
the dark-shaded region shows the contribution from
cosmic variance.
\label{ClFig}
\end{figure}

\clearpage
\begin{figure}[phbt]
\centerline{\epsfxsize=16cm\epsfbox{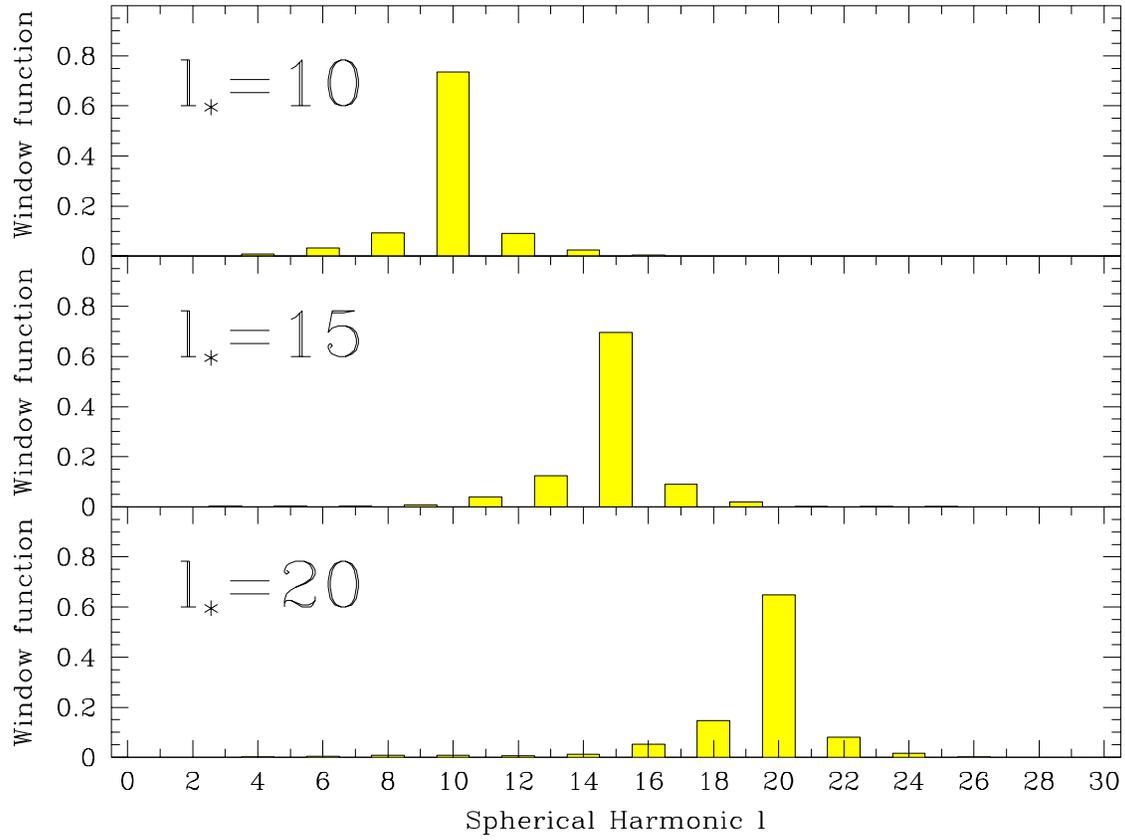}}
\caption{
How the window functions depend on the target multipole.
}
\label{LocationFig}
\end{figure}

\clearpage
\begin{figure}[phbt]
\vskip-2.5cm
\centerline{\epsfxsize=16cm\epsfbox{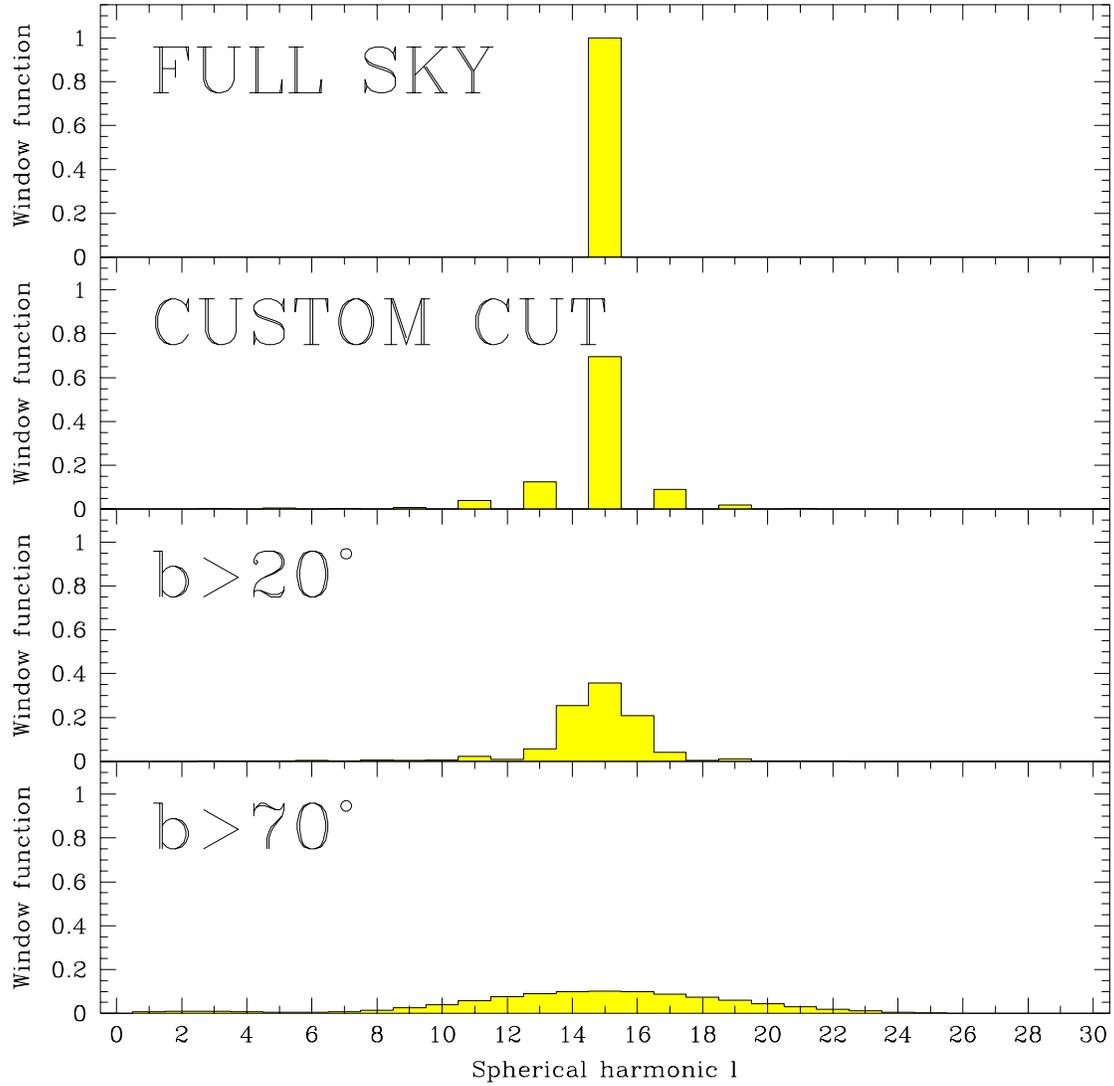}}
\caption{
How the window functions depend on the sky coverage.
}
As the sky coverage decreases, the window function
widens from a Kronecker delta to have a width
$\Delta\l\sim 1/\Delta\theta$. 
In addition, since the custom cut is almost symmetric, it
approximately preserves the orthogonality of
even and odd multipoles.
\label{WidthFig}
\end{figure}

\clearpage
\begin{figure}[phbt]
\centerline{\epsfxsize=14cm\epsfbox{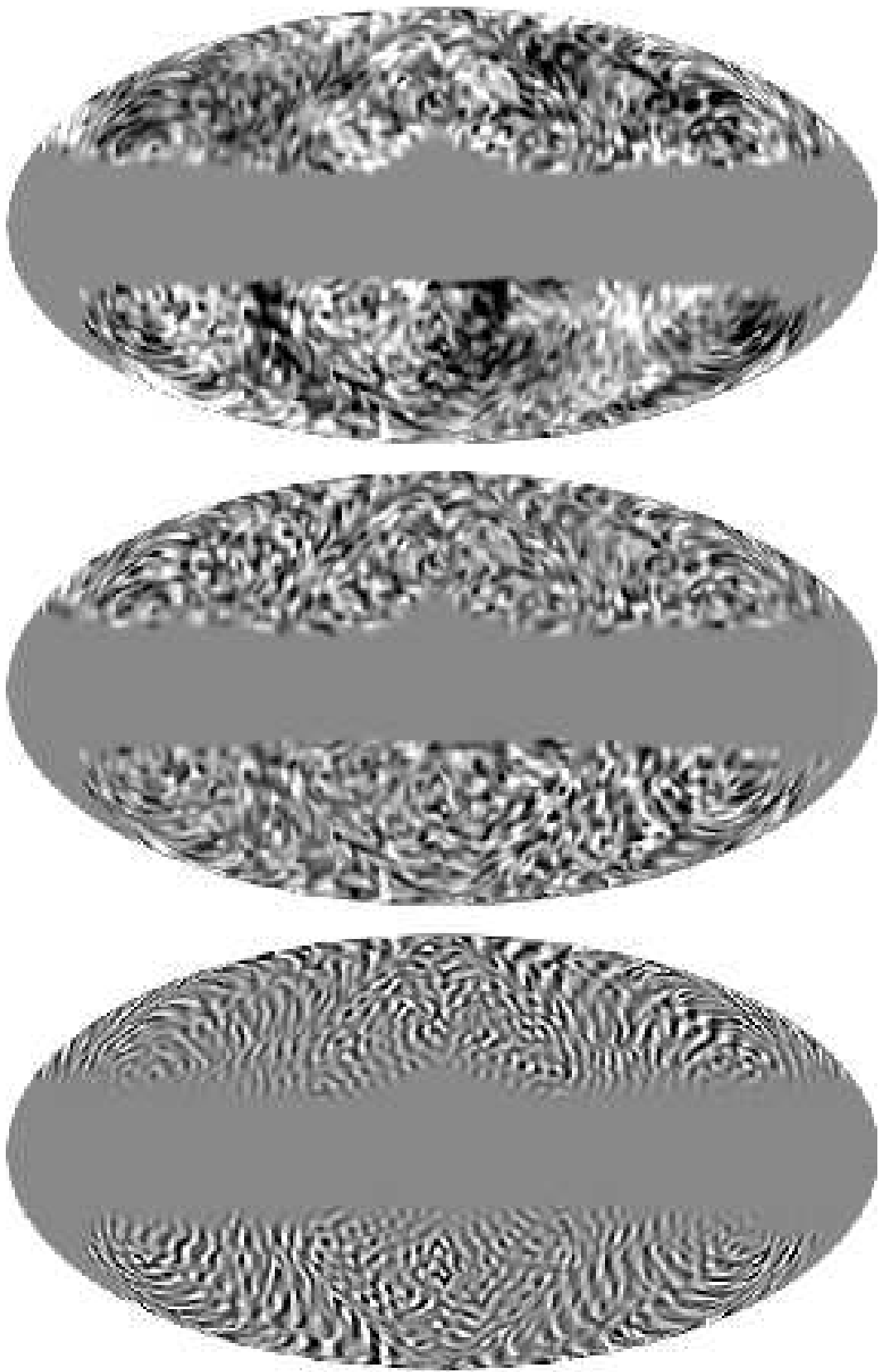}}
\caption{
Sky maps before and after power tilting.
}
The 4-year COBE/DMR data (top) is shown after multiplication 
by the inverse pixel covariance matrix corresponding to the 
actual noise level (middle) and 
corresponding to the noise-level projected for {\it Planck} (bottom).
\label{MapsFig}
\end{figure}

\clearpage
\begin{figure}[phbt]
\centerline{{\vbox{\epsfysize=14cm\epsfysize=16cm\epsfbox{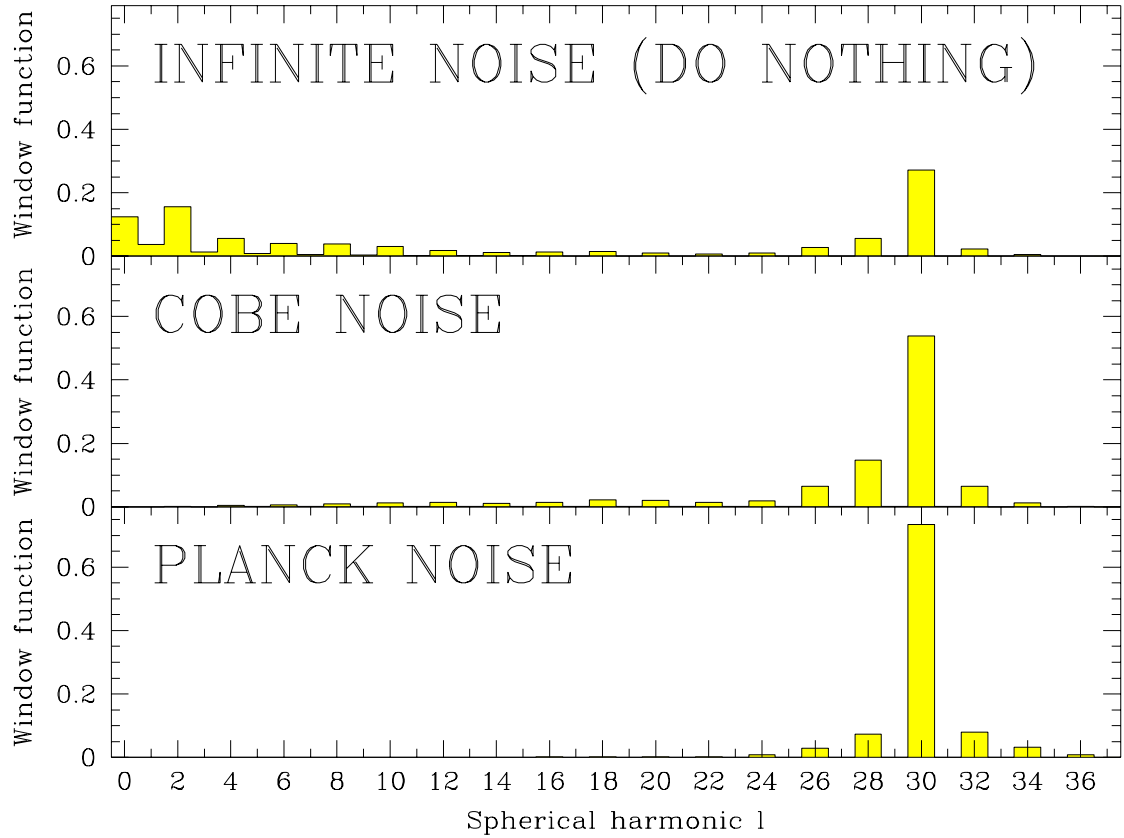}}}}
\caption{
Window functions before and after power tilting.
}
Same as the previous figure, but in the Fourier (multipole) domain,  
showing the corresponding window functions for $\l_*=30$.
\label{TiltFig}
\end{figure}

\clearpage
\begin{figure}[phbt]
\centerline{{\vbox{\epsfxsize=16cm\epsfbox{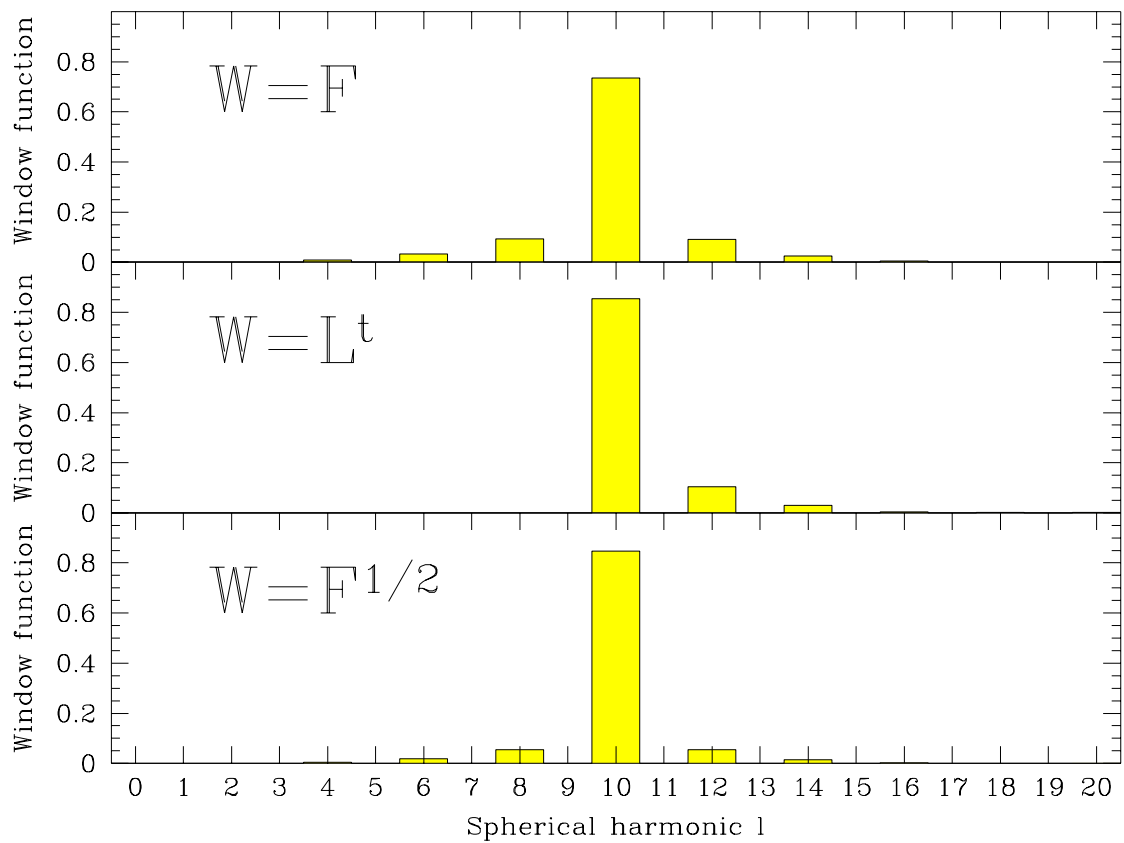}}}}
\caption{
The $\l_*=10$ window functions for three of the data-presentation
methods discussed, corresponding to using the original $\y$, 
Cholesky decomposing the Fisher matrix ($\F=\L\L^t$)
and taking its square root, respectively.
}
\label{UncorrWindowFig}
\end{figure}

\clearpage
\begin{figure}[phbt]
\centerline{{\vbox{\epsfxsize=16cm\epsfbox{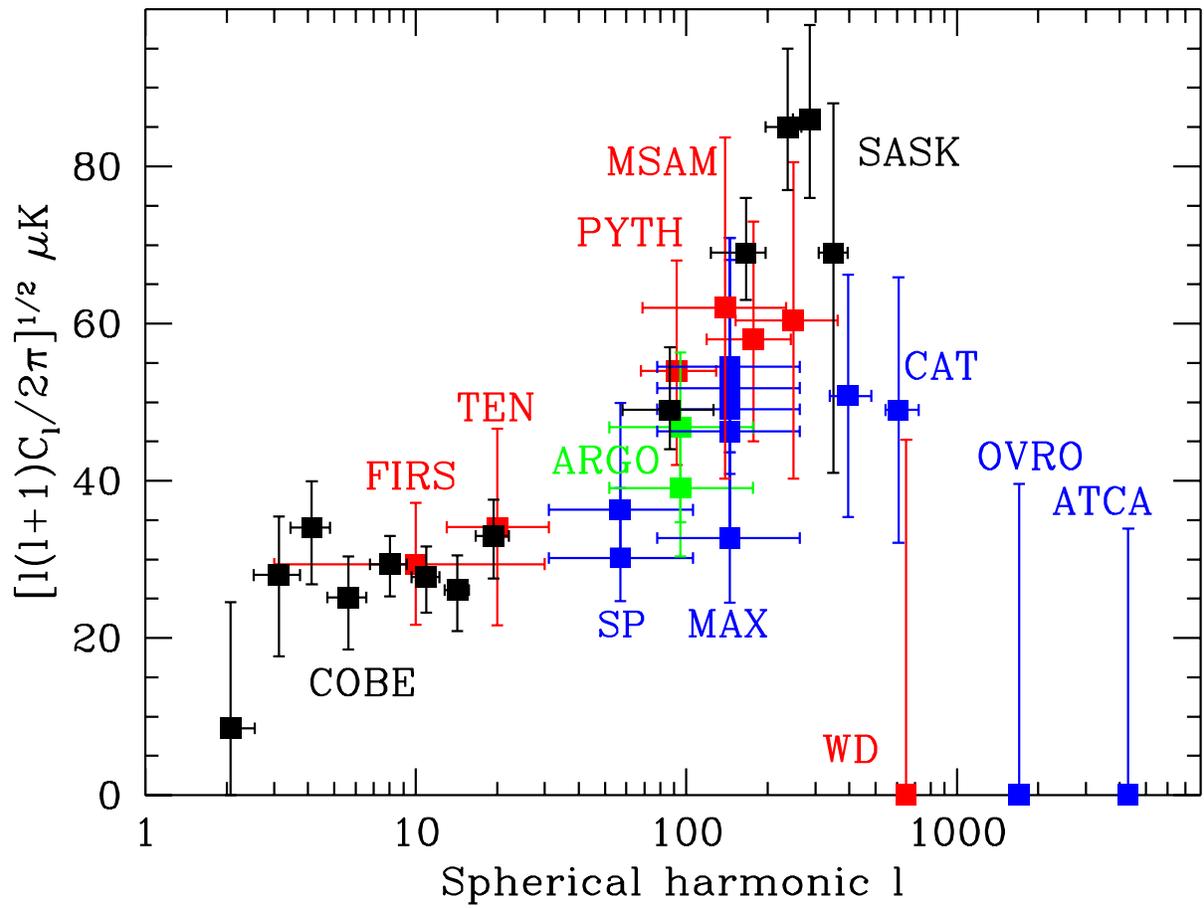}}}}
\vskip1cm
\caption{
The power spectrum observed by COBE/DMR binned into 8 bands and compared 
with other experiments. 
}
\label{ExperimentsFig}
\end{figure}


\end{document}